\documentclass[proof]{WileyASNA-v1}
\articletype{Article Type}
\usepackage{stix}
\received{XXXX}
\revised{XXXX}
\accepted{XXXX}

\begin{document}
%

\title{Confirmation of the detection of B-modes in the Planck polarization maps}

\author[1]{H. U. N{\o}rgaard - Nielsen}

\authormark{AUTHOR ONE}
\address[1]{\orgdiv{National Space Institute (DTU Space)},
    \orgname{Technical University of Denmark}}

\corres{H. U. N{\o}rgaard - Nielsen, \email{hunn@space.dtu.dk}}

\presentaddress{Elektrovej, DK-2800 Kgs. Lyngby, \country{Denmark}}



\abstract{One of the main problems for extracting the Cosmic Microwave Background (CMB) from submm/mm observations is to correct for the Galactic components, mainly synchrotron, free - free and thermal dust emission with the required accuracy. Through a series of papers, it has been demonstrated that this task can be fulfilled by means of simple neural networks with high confidence. The main purpose of this paper is to demonstrate that the CMB BB power spectrum detected in the Planck 2015 polarization maps is present in the improved Planck 2017 maps with higher signal-to-noise ratio. Two features have been detected in the EB power spectrum in the new data set, both with S/N $\sim$4 . The origin of these features is most likely leakage from E to B with a level of about 1 per cent. This leakage gives no significant contribution to the detected BB power spectrum. The TB power spectrum is consistent with a zero signal. Altogether, the BB power spectrum is not consistent with the 'canonical' tensor-to-scalar models combined with gravitational lensing spectra. These results will give additional strong arguments for support to  the proposed polarization satellite projects to follow up on the Planck mission .}


\jnlcitation{\cname{\author{N{\o}rgaard - Nielsen, H. U.} {\cyear{2017}},
\ctitle{Confirmation of the detection of B-modes in the Planck polarization maps}, \cjournal{AN}, \cvol{XXX}}.}

\maketitle


\section{Introduction}

Since the discovery of the Cosmic Microwave Background radiation (CMB) by Penzias and Wilson(1965), investigations of this radiation has been a central feature in order to get new information about the earliest evolution of the Universe. Due to technical limitations, until recently most of the observational effort has been concentrated to obtain very accurate temperature measurements. Rees (1968) showed that polarization observations of CMB on larger scales could be a very important way to study the very early history of the Universe. Symmetries in the production and growth of the polarization signal are constraining the configurations  of the CMB polarization. Density (scalar) perturbations produce temperature (T) fluctuations and E (curl - free) polarization modes, while tensor (gravitational waves) produce both T, E and B (divergence  free) modes.

It is generally assumed that the initial inhomogeneities in the Universe were Gaussian distributed. Linear theory predicts that the CMB fluctuations are also Gaussian, and that the CMB spectrum can be fully described by 4 power spectra TT, EE, BB and TE while the TB and EB power spectra should zero due to parity constraints (e.g. Kamionkowski et. (1997).

CMB polarization measurements were provided by the WMAP satellite, the final results given in Bennett et al. (2013). For r, the tensor-to-scalar ratio, Bennett et al. found, for WMAP-only data, an upper limit of 0.38.

Ade et al. (2014) report a detection of B-modes in a 400 sq. degree area on the sky with the BICEP2 instruments, observing from the South Pole.
These observations were done at only one frequency, 150 GHz, implying that their estimate of polarized Galactic emission in this small area on the sky is uncertain.

Ade et al. (2015) attacked this problem by combining BICEP2 + Keck 150 GHz data and Planck 30 GHz-353 GHz observations in this area.  This investigation determined the amplitude of the lensing  spectrum 1.12 $\pm$ 0.18, relative to the standard $\lambda$CDM model, and an upper limit on the tensor-to-scalar ratio r $<$ 0.13, with 95 percent confidence.

The Planck Collaboration XV (2016) determined the CMB lensing potential at a level of 40$\sigma$, while the Planck Collaboration XIII (2016) found an upper limit of 0.11 on the tensor-to-scalar ratio.

For the Planck Collaboration, a key issue has always been to carefully investigate the data flows from the detectors and find ways to correct for systematic errors. Since the release of the Planck data in 2015, this effort has been continued and significant improvement in the final Planck frequency maps has been obtained. Since no new observations have been obtained between 2015 and 2017 , the improvements are mainly due to removal of remaining systematic errors.

In a series of papers, the capabilities of neural networks for dealing with mm/submm observations have been investigated by N{\o}rgaard - Nielsen \& J{\o}rgensen (2008),  N{\o}rgaard - Nielsen \& Hebert (2009),
N{\o}rgaard - Nielsen (2010), and N{\o}rgaard - Nielsen (2012).
The last two papers showed, by using data from the WMAP satellite, that simple neural networks can extract the CMB temperature and polarization signals with excellent accuracy.

In N{\o}rgaard - Nielsen (2016, hereafter NN1) the Planck polarization frequency maps, released in 2015, the Stoke Q and U parameters were extracted by means of simple neural networks. The BB power spectrum was detected in 100 $\leq$ l $\leq$ 275 with S/N = 4.5. It was demonstrated that the contribution from Galactic emission and remaining systematic errors in the maps were very small, but they could not be completely ruled out. Due to the improvement in the Planck 2017 polarizations maps, it is feasible to repeat the 2016 analysis. As in NN1, the this paper is concentrated on the reliability of the detected polarization power spectra.

The structure of the paper is the following: Sect. 2 and 3 give a short overview of the different analysis tools applied in NN1, the power spectra from the extracted Q and U maps are presented in Sect. 4, while the fully calibrated power spectra are shown in Sect. 5, the contamination of the power spectra is discussed in Sect. 6. Conclusions are given in Sect.7 .

\section{The final Planck polarization maps}
In the present analysis, the final Planck frequency maps (named DX12) have been exploited:
\begin{itemize}
\item{only the final Q and U maps in the frequency range 30 GHz - 353 GHz has been exploited}
\item{no attempt has been performed to correct the Planck frequency maps to the same point spread function}
\item{the analysis has been performed exclusively on the Planck frequency maps, implying that no auxiliary data or information has been exploited}
\end{itemize}

  From the final set of Planck frequency maps, 4 sets have been exploited in this paper: the 2 half mission maps (HM1 and HM2) and the maps combined from the ODD years and the EVEN years of the mission. The precise definition of this data sets can be found in Planck Collaboration I (2017)

\section{The road from frequency maps to CMB power spectra}
In this paper, the procedure for extracting the Stoke parameters Q and U from the final Planck frequency maps follows closely the procedure used in NN1. In this section, the key features of this procedure are outlined.

\subsection{The applied foreground model}
As for the previous investigations, the main purpose is only to extract the CMB signal. Therefore, meaningful  physical models of the Galactic foreground components are not required, only a coherent mathematical model of the non - CMB signal (all signals that are not related to the CMB) is needed.

The Independent Component Analysis (Hyvarinen et al. 2001, ICA) has turned out to be well suited to provide a simple model of the non - CMB signal. Briefly, ICA is a mathematical method to extract a set of independent components, in this case from the 7 Planck frequency maps.
Like in NN1, each of the HM1, HM2, ODD and EVEN data sets are split into independent components by ICA. By analysing these component maps, it is evident that 2 components contain nearly all the extended Galactic emission. For each of these components, ICA provides a normalized spectrum and an amplitude per sky pixel, which are used in setting up the neural networks, see Sect. 3.2.

\subsection{The neural network method}
In the series of papers referred above, it has been demonstrated that neural networks are a very effective tool for extracted CMB from mm/submm observations. The setup of the neural networks is the same as previously:
\begin{itemize}
\item{simple multi layer perceptron neural networks with 2 hidden layers have been exploited}
\item{the input for the neural networks are seven inputs channels (30 GHz - 353 GHz) and one output channel (Q or U), all in units of $\mu$K}
\item{the standard non - linear activation function 'tansig' has been exploited}
\end{itemize}

The scheme for setting up the data for training the neural network is:

\begin{enumerate}
\item{defining the required intervals of the amplitudes of the 2 ICA components and for Q or U}
\item{draw 3 random numbers within these intervals}
\item{calculate the combined spectrum}
\item{for each input channel, add random Gaussian noise within the range corresponding to the Planck hit maps in the different data sets}
\item{run these steps N(train) times, to get reasonable number of spectra to train the network}
\item{train the network to establish the transformation between the input channels and the output channel}
\item{test the accuracy of the network by running an independent TEST data set, with N(test) spectra obtained in the same way as the TRAIN data set, through the network.}
\item{if the accuracy is satisfactory (meaning the systematic errors as function of the input parameters are very small and the residuals are Gaussian distributed), the Planck data set is run through the network}
\end{enumerate}

The main advantages of using neural networks are:

\begin{itemize}
\item{the number of weights needed to set up the full transformation between the 7 input channels and 1 output channel (Q or U) is in our case $~$ 80, small compared to the number of pixels in each of the Planck frequency maps ($\sim$ 12.000.000)}
    \item{the foreground model is only using the input frequency maps themselves (neither auxiliary data nor assumptions about e.g. the spectral behaviour of the different foreground components are applied)}

\item{a small number of parameters are needed in the foreground model, since a physical meaningful model is not required }

\item{the network is set up to extract signals with the same spectral behaviour as CMB}

\item{the available frequency range is exploited}

\end{itemize}

 \begin{figure}[ht]
\centering
\includegraphics[width= 3.0in]{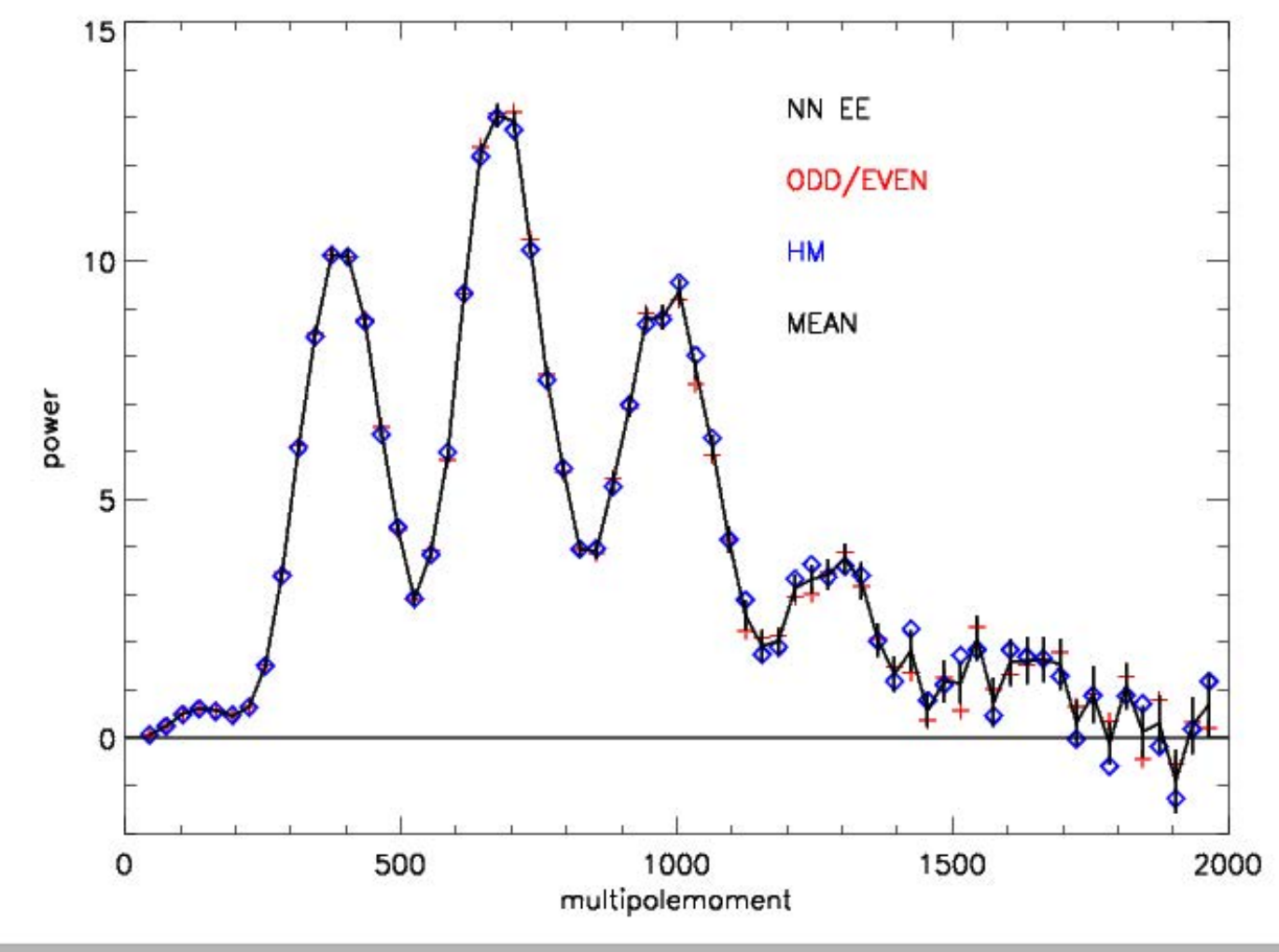}
\caption{The CC HM and OE EE power spectra together with the mean. It is seen that these spectra fit nicely together up to l $=$ 1700. Y - axis: $l(l +1)/(2\pi) C^{EE}_{l}[\mu K^{2}]$}
\label{fig_1}
\end{figure}

\subsection{Applied sky mask}
 In order to avoid residual contamination by emission from the MIlky Way in the extracted power spectra, the apodized CAMSPEC mask used in NN1, covering about 63 per cent of the sky, has also been applied here. Since the contribution from point sources is expected to be insignificant, it is also neglected in this work.

\subsection{The Planck simulations}

In the process of establishing  the Planck 2015 data, the whole Planck data chain has been extensively simulated (FFP8). In NN1, this data set was mainly used for calibrating the derived power spectra to an absolute scale. The basic simulations behind FFP8 has been improved, but from a calibration  point of view, the improvements are not significant. Therefore, the NN1 calibration scheme has also been followed here.

\begin{table}[h]
\caption{Definition of the 'standard' Planck high pass filter, defined with $l_{1}$ = 20  and $l_{2}$ = 40 (Planck 2015 IX)}
\centering
\begin{tabular}{c c}
l range & value\\
\hline
$l < l_{1}$ & 0.0\\
$l_{1} < l < l_{2}$ & $0.5[1~-~cos(\pi\frac{l~-~l_{1}}{l_{2}~-~l_{1}})]$ \\
$l > l_{2}$& 1.0\\
\hline
\label{high_pass}
\end{tabular}
\end{table}

\subsection{The high pass filter}

It turned out that special treatments of the Planck 2015 maps at multipoles less than 20 were needed, mainly due to the uncertainty in the contribution of large scale systematic errors still in the frequency maps. Although there have been significant improvements in the frequency maps since 2015, the discussion of this range of multipole moments is outside the goal of this investigation, and the 'standard' Planck high pass filter has been introduced (Table 1) as in NN1.

\section{The power spectra of the observed Planck polarization maps}
The accuracy of cross correlation (CC) power spectra is superior compared to auto correlation (AC) power spectra. Therefore, the discussions in this paper are concentrated on the HM1 $<$-$>$ HM2 (HM) and the ODD $<$-$>$ EVEN (OE) CC power spectra.

All power spectra analysed in this paper have been extracted with the HEALPix \textbf{anafast} IDL routine (\citet{Gors05}).

The l - interval of the power spectra are the same as for the Planck 2015 EE power spectrum.

\begin{figure}[!]
\centering
\includegraphics[width=3.0 in]{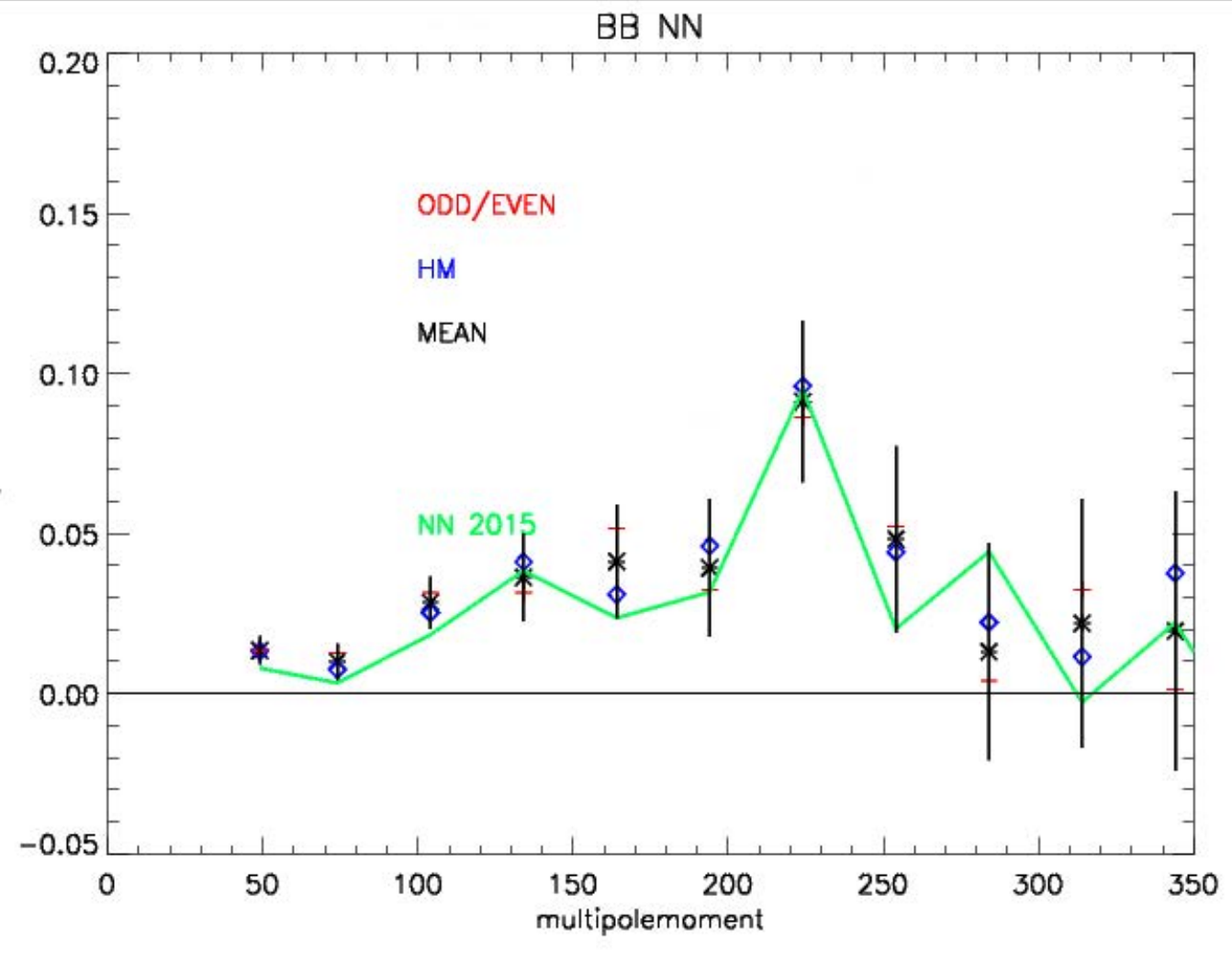}
\caption{The CC HM and OE BB power spectra, compared with the 2015 BB power spectrum (NN1). Taking the uncertainty into account, a reasonable agreement is seen between the $<$2015$>$ and $<$2017$>$ spectra. Y - axis: $l(l +1)/(2\pi) C^{BB}_{l}[\mu K^{2}]$}
\label{fig_2}
\end{figure}

\subsection{The EE power spectra}

The CC HM and EO EE power spectra together with the mean are shown in Fig. \ref{fig_1}. It is seen that the HM and OE power spectra are very similar up to l $\sim$ 1700.

\subsection{The BB power spectra}

The derived CC HM and OE BB power spectra and the mean are shown in Fig. \ref{fig_2}, together with the CC BB power spectra from NN1. Taking the accuracy of these power spectra into account a reasonable agreement is seen. It is  seen that the main features in the 2015 spectrum (NN1) is still present.

\begin{figure}[!]
\centering
\includegraphics[width=2.5 in]{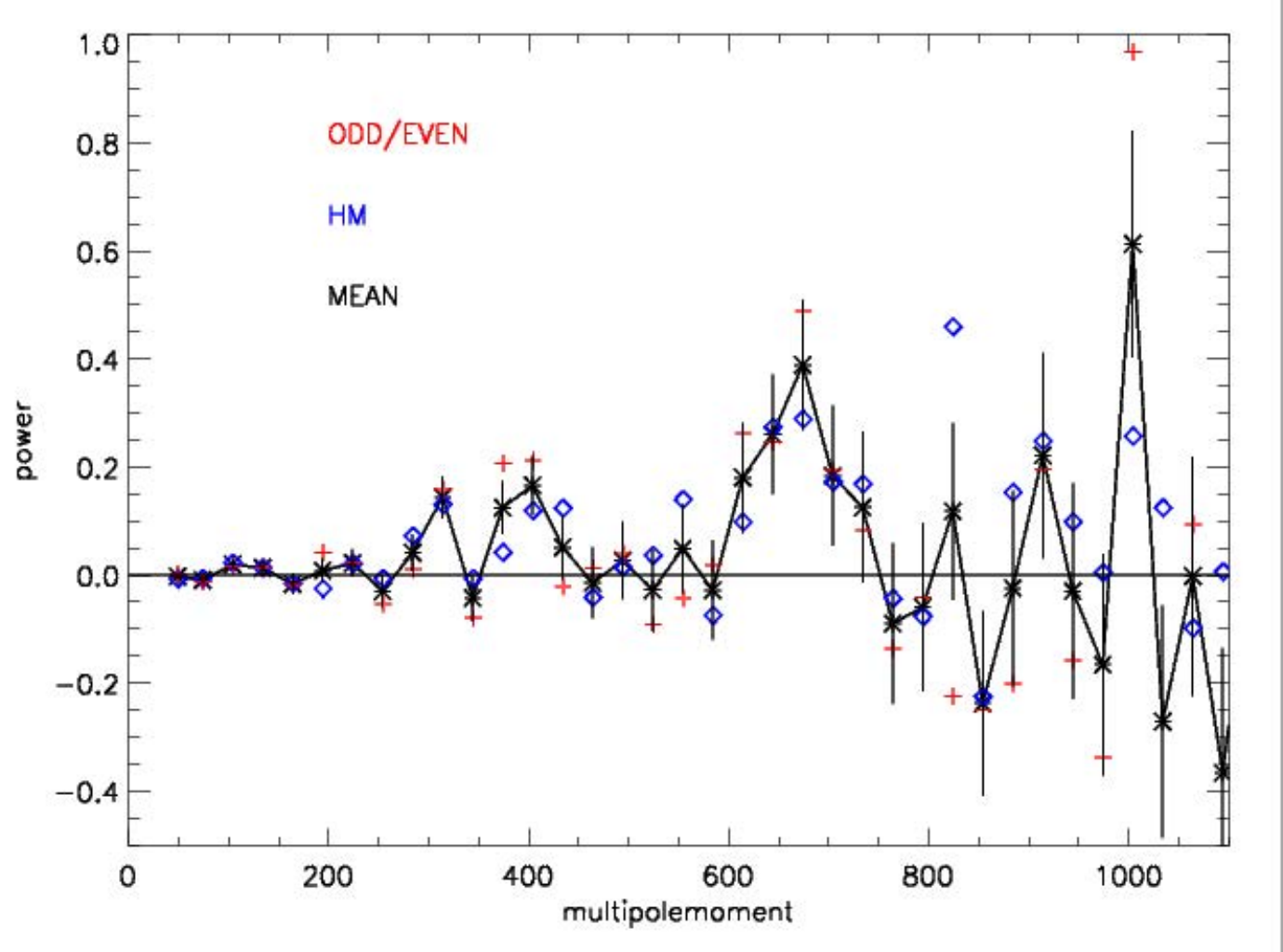}
\caption{The CC EB power spectrum. Y - axis: $l(l +1)/(2\pi) C^{EB}_{l}[\mu K^{2}]$}
\label{fig_3}
\end{figure}

\subsection{The EB power spectrum}

The CC EB power spectrum shows 2 features around l = 675 and l = 375, each detected with a S/N $\sim$4 (Fig.\ref{fig_3}). The positions of these feaures correpond to the postions of the first 2 peaks in the EE power spectrum (Fig.\ref{fig_1}).

\begin{figure}[!]
\centering
\includegraphics[width=3.1 in]{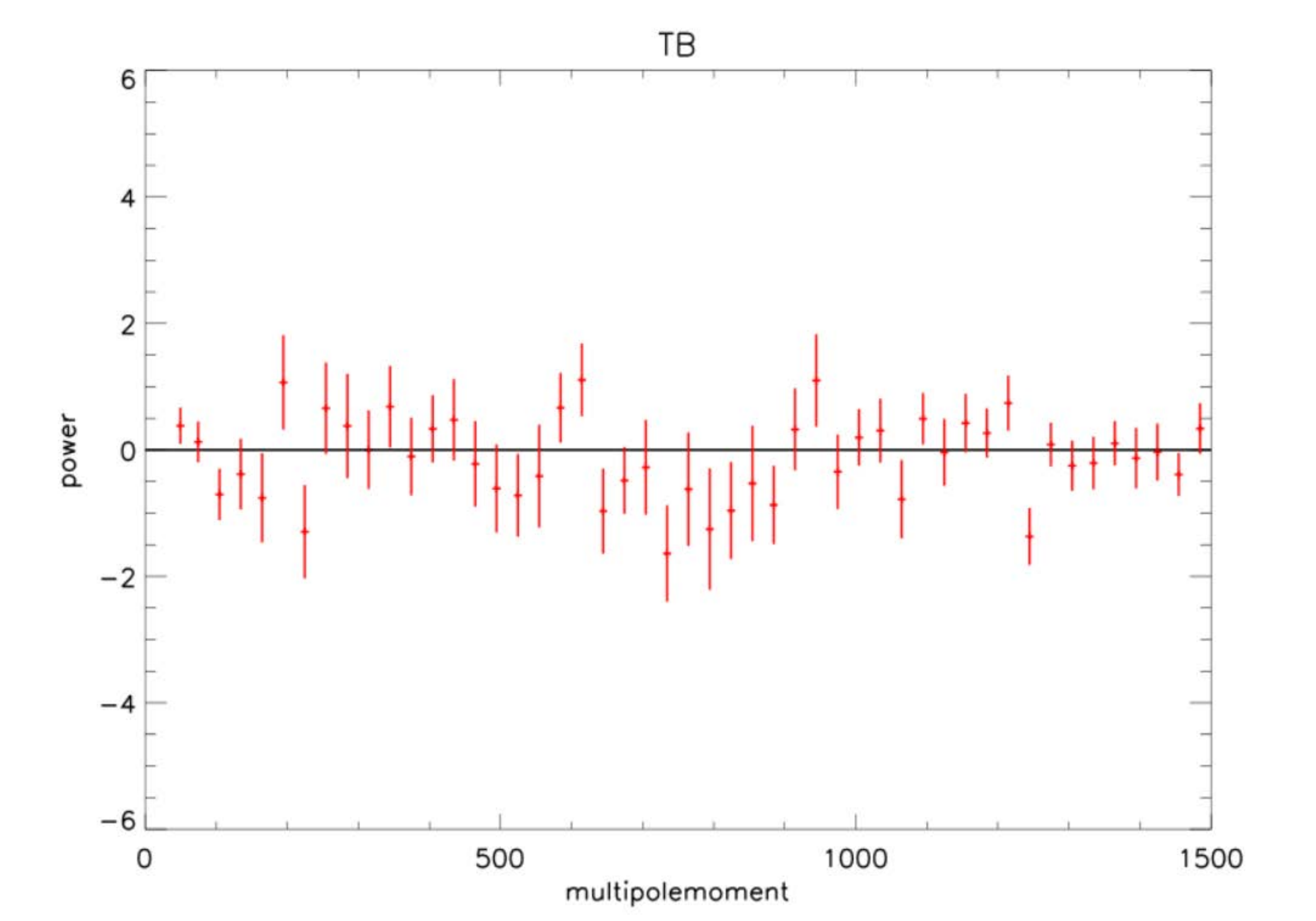}
\caption{The CC TB power spectrum, it is evident that this spectrum is consistent with a zero signal. Y - axis: $l(l +1)/(2\pi) C^{TB}_{l}[\mu K^{2}]$}
\label{fig_4}
\end{figure}

\subsection{The TB power spectrum}

The CC HM TB spectrum is presented in Fig. \ref{fig_4}. It is clear that no part of this spectrum has been significantly detected. For 100 $\leq$ l $\leq$ 750 the average power is -0.14 $\pm$ 0.13 ,

\begin{figure}[!]
\centering
\includegraphics[width=3.0 in]{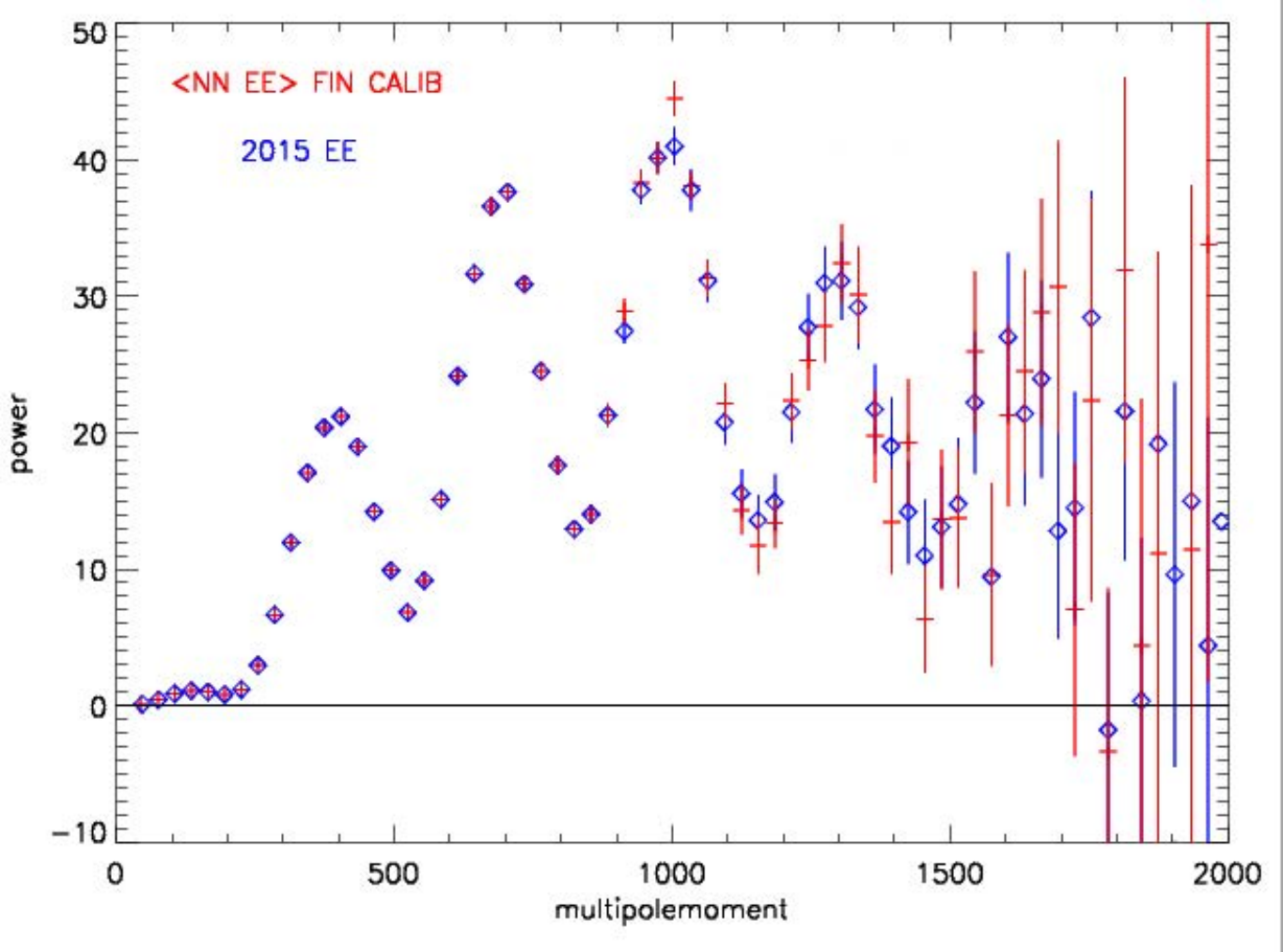}
\caption{The fully calibrated CC $<$HM, OE$>$  EE power spectrum, compared with the Planck 2015 EE power spectrum (Planck 2015 XI) including error bars. An excellent agreement is seen for l $<$ 1650. Y - axis: $l(l +1)/(2\pi) C^{EE}_{l}[\mu K^{2}]$}
\label{fig_5}
\end{figure}

\section{The calibration of the power spectra}

The basic calibration scheme used in NN1 (mainly corrections for the sky mask, the sensitivity of the Planck detectors and for the effective point spread function of the NN method) has been applied. Briefly, these corrections were estimated by combining a set of simulated FFP8 frequency maps with realistic noise maps and theoretical tensor-to-scalar maps, calculated by the CAMB software package (see http://CAMB.info).

\section{The fully calibrated EE, BB and EB power spectra}

The mean, fully calibrated, CC $<$HM, OE$>$ EE spectrum are given in Fig. \ref{fig_5} together with the Planck 2015 EE power spectrum, a good agreement is seen. As in NN1, the error bars given in the figures are extracted from the scatter around the spectra in each l - interval.

\begin{figure}[!]
\centering
\includegraphics[width=3.0 in]{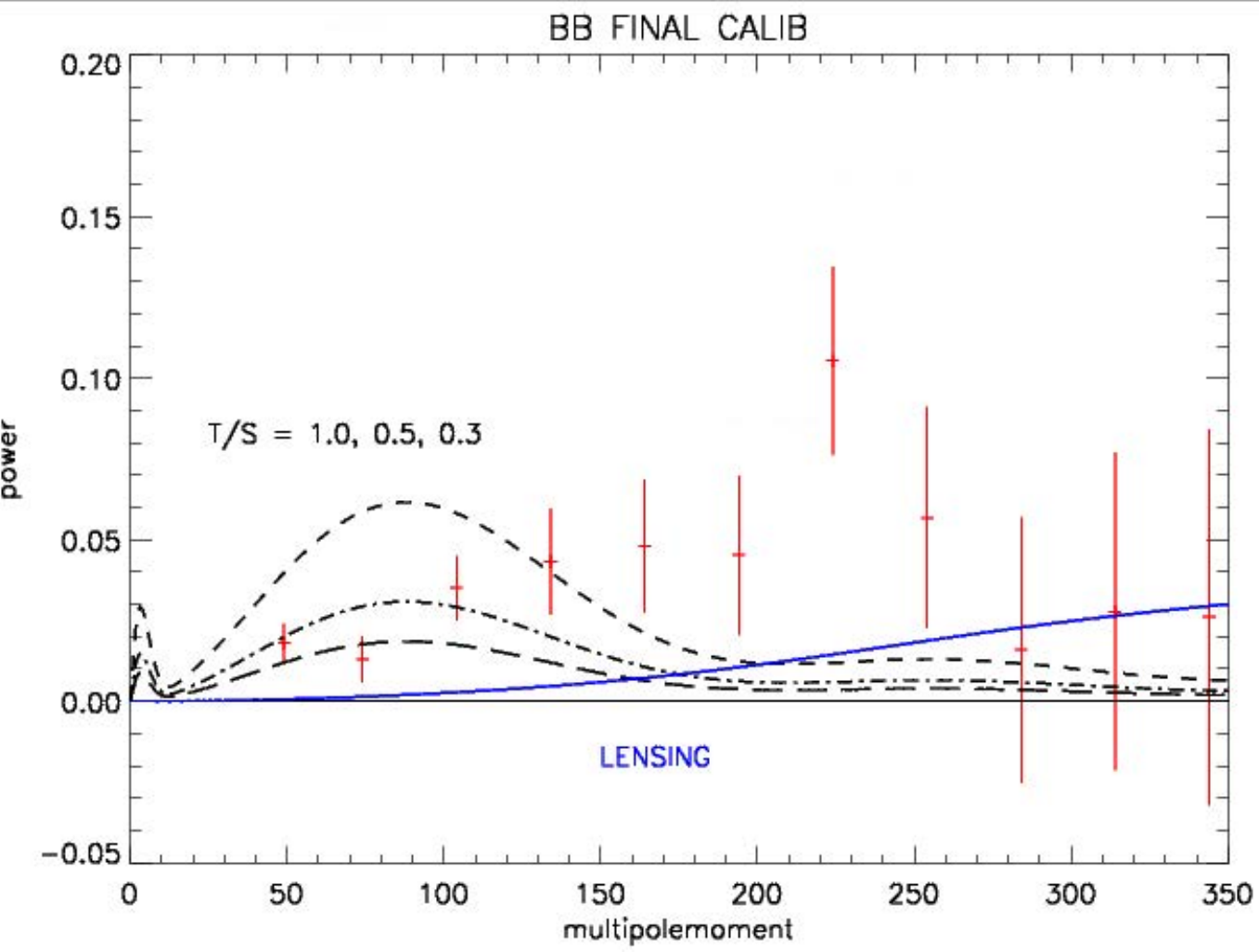}
\caption{The fully calibrated CC $<$HM, OE$>$ BB power spectrum, compared with PLANCK 2015 estimates of the BB power spectrum.  CAMB BB tensor spectra with tensor to scalar ratios of 0.5 and 0.3 are also shown. Y - axis: $l(l +1)/(2\pi) C^{BB}_{l}[\mu K^{2}]$}
\label{fig_6}
\end{figure}

In Fig. \ref{fig_6}, the mean, fully calibrated CC  $<$HM, OE$>$ BB power spectrum is shown. The excess from l = 100 to l = 275, discussed in NN1, are confirmed with improved S/N = 6.6 (compared to 4.5 in NN1 ), the feature 175 $\leq$ l $\leq$ 275 has a S/N of 4.1 . The theoretical CAMB spectra with tensor to scalar ratios of 1.0, 0.5, 0.3 are also shown together with the CAMB lensing spectrum. It is clear that a combination of these CAMB spectra will have difficulties in fitting the observed spectrum.

\begin{figure}[!]
\centering
\includegraphics[width=3.0 in]{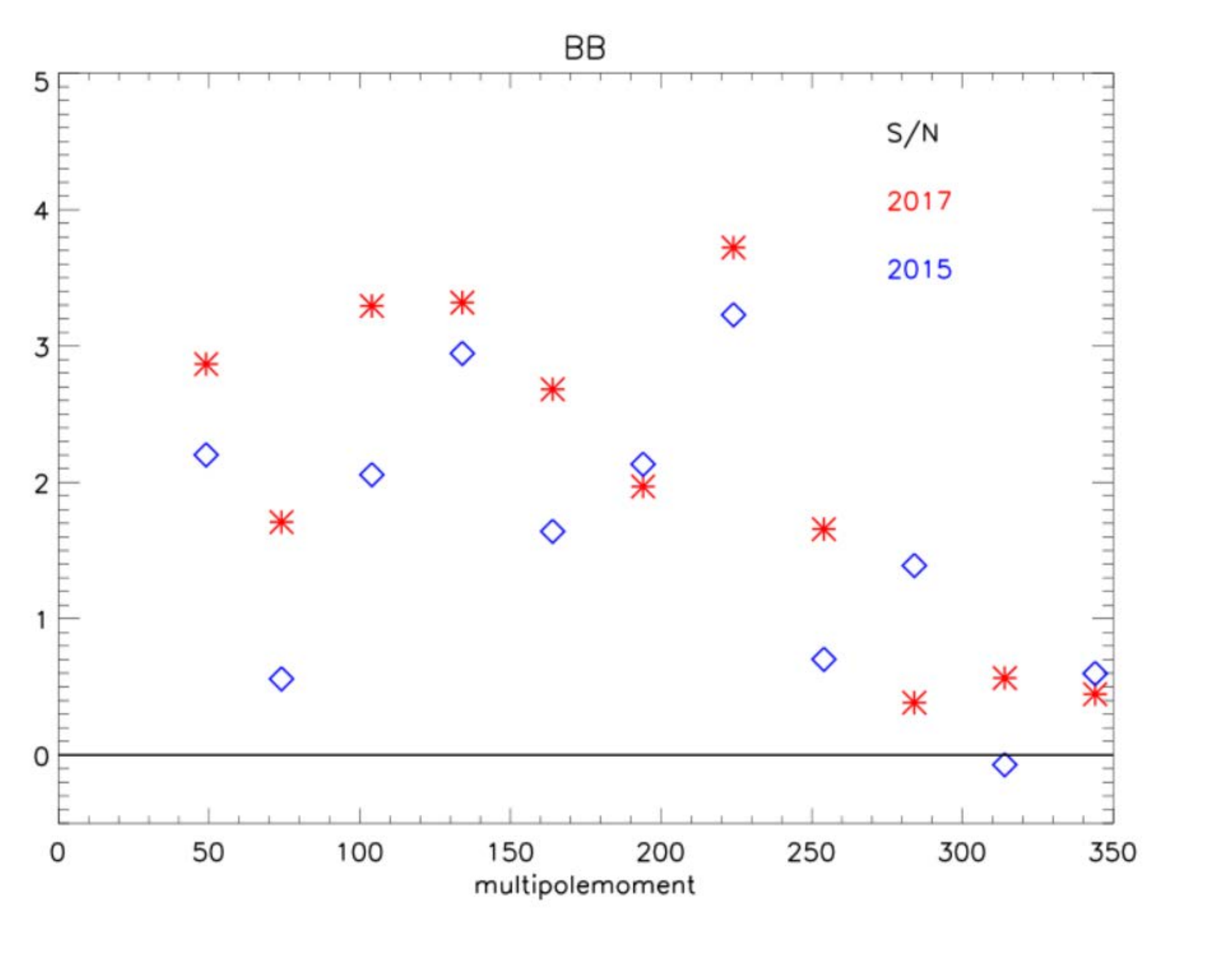}
\caption{The Signal to Noise of the 2017 BB power spectrum, compared with previous obtain spectrum. A considerable improvement is evident}
\label{fig_7}
\end{figure}

In Fig. \ref{fig_7}, the difference in S/N between the BB power spectrum found in  NN1 and the spectrum given here is highlighted. The improvement is evident.

\begin{figure}[!]
\centering
\includegraphics[width=2.5 in]{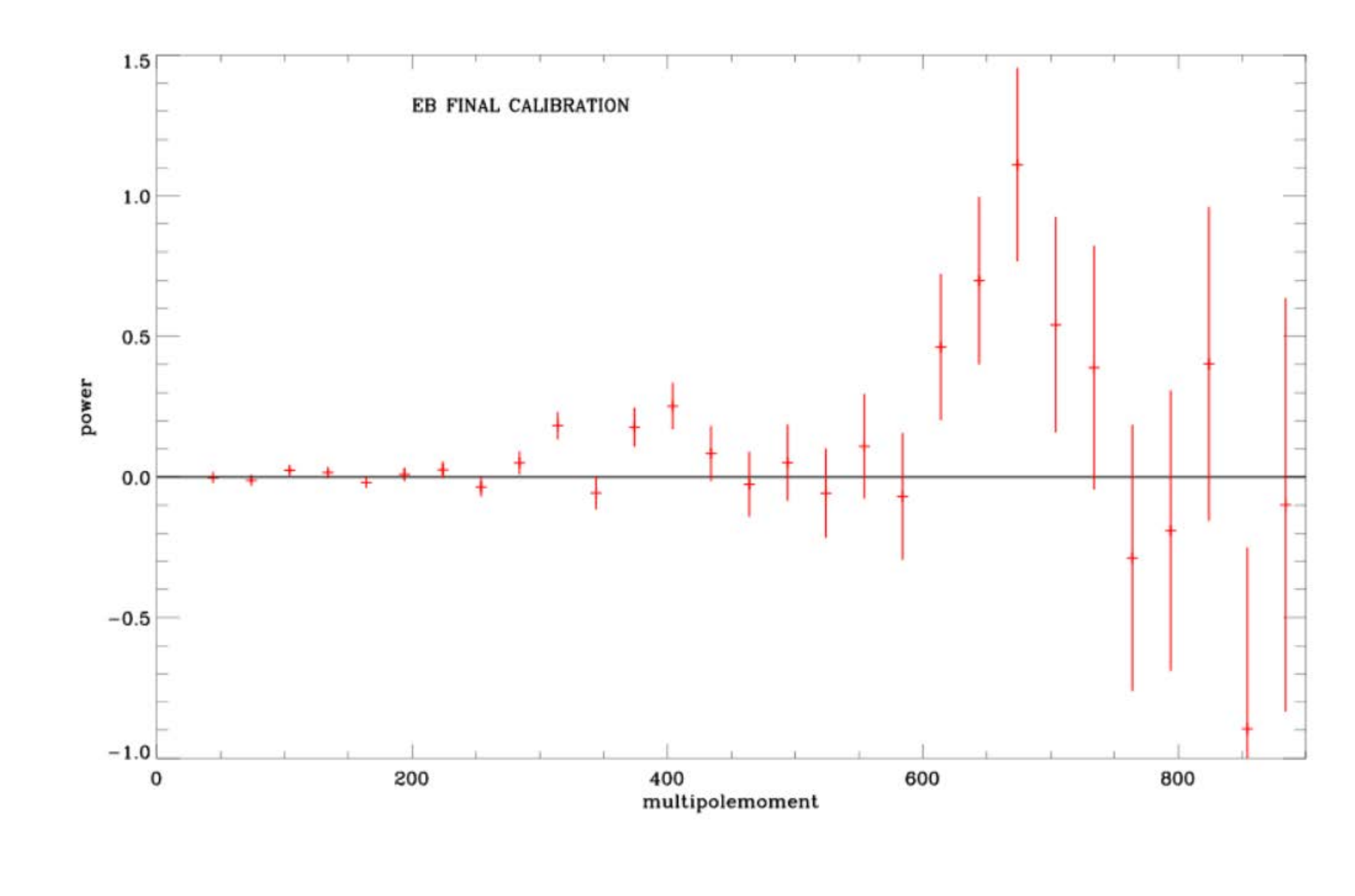}
\caption{The fully calibrated CC $<$HM, OE$>$ EB power spectrum. Y - axis: $l(l +1)/(2\pi) C^{EB}_{l}[\mu K^{2}]$}
\label{fig_8}
\end{figure}

The fully calibrated CC EB power spectrum is shown in Fig. \ref{fig_8}. The 2 features around l $=$ 675 and l $=$ 375 have $S/N$ $=$ 4.4, and 4.2, respectively. These features could indicate a violation of the parity assumption, but the coincidence with the positions with the first 2 peaks of the EE power spectrum could also indicate a leakage from E to B. To explain the strength of the EB features, a leakage of 1  - 2 per cent is sufficient. At this stage, it is difficult to exclude a leakage of this level. Such a leakage would give a insignificant contribution to the BB power spectrum compared to the noise level at the positions of the features. In Fig. \ref{fig_1}, it is seen that the EE power is much weaker than the first peaks in the l - range relevant for the BB spectrum discussed in this paper, implying that the E to B leakage gives no significant contribution to the detected BB power spectrum.

\section{Remaining systematics effects from the reduction of the Planck frequency maps }
As emphasized above, important improvements in removing residual systematics have been obtained by the LFI and HFI instrument teams since 2015.

As emphasized in the two Planck 2015 detector papers (Planck Collaboration III and VIII (2016)) the systematic errors on larger scales are the most difficult to remove, but the high-pass-filter (Table \ref{high_pass}) is designed to remove this kind of systematics.

Planck Collaboration III (2017) gives a detailed summary of all the systematic errors found during the reduction phase of the HFI data and limits on their amplitudes. From their Table 7 it is seen that main contributor of remaining systematic errors in the frequency maps is the uncertainties in the ADC linear corrections. For the 100 GHz, 143 GHz and 217 GHz channels (the most sensitive channels to the CMB signal on Planck) the residual level of systematic errors is 1 - 3 $10^{-3}$ $\mu \textrm{K}^{2}$, peaking at small l's, implying that they give only minimal contributions to the cross power spectra presented here.

For the other important contribution to systematics e.g. calibration and bandpass mismatch are nearly an order of magnitude lower. Further, the neural networks are design to extract components with a flat frequency spectrum. Altogether, it is reasonable to assume that the observed BB and EB cross spectra are not polluted significantly by residual systematic errors.

The E/B leakage due the uncertainties in the beams of the Planck optical system is estimated to be less than $3 x 10^{-3} \mu K^{2}$, which is small compared to the detected BB power spectrum (see Fig. 6).

\section{The E/B leakage}

Due to the fact that most, if not all, CMB experiments extract the power spectra from only a fraction of the sky (due to the strong signal from the MIlky Way in the relevant frequency range) the E modes will to some extend leak into the B modes. This problem has been discussed in a number of papers e.g. Kendrick (2006) and Grain et al. (2009). They show the problem can be significantly reduced by a proper apodization of the mask. Most of these investigations have analysed the problem if small areas on the sky, say 1 per cent, are observed, relevant for e.g. balloon flights. Since the Planck Mission is covering the whole sky, it is possible to analyse a large part of the sky, it is only necessary to exclude the brightest area of the Milky Way. As emphasized in Section 3.3, data from about 60 percent of the sky has been exploited in this paper.
To investigate the level of leakage in the detected BB power spectrum, the BB power spectra has been calculated from the CAMB T/S = 1.0 models with different masks applied. The leakage (defined as the difference of the power spectrum included a mask minus the full sky power spectrum). The masks applied are similar to the mask described in Section 3.3, except for the size of the sky coverage.
In Fig. 9, it is seen that the level of the E/B leakage for the mask used in calculated the BB power spectrum (the purple line) is insignificant compared to the detected spectrum(Fig. 6).

\section{Contamination by Galactic emission}
As emphasized above, the neural networks are setup to extract a signal corresponding to a flat spectrum, but at low flux levels the noise of the observations presents a challenge for the neural networks.

\begin{figure}[!]
\centering
\includegraphics[width=3.5 in]{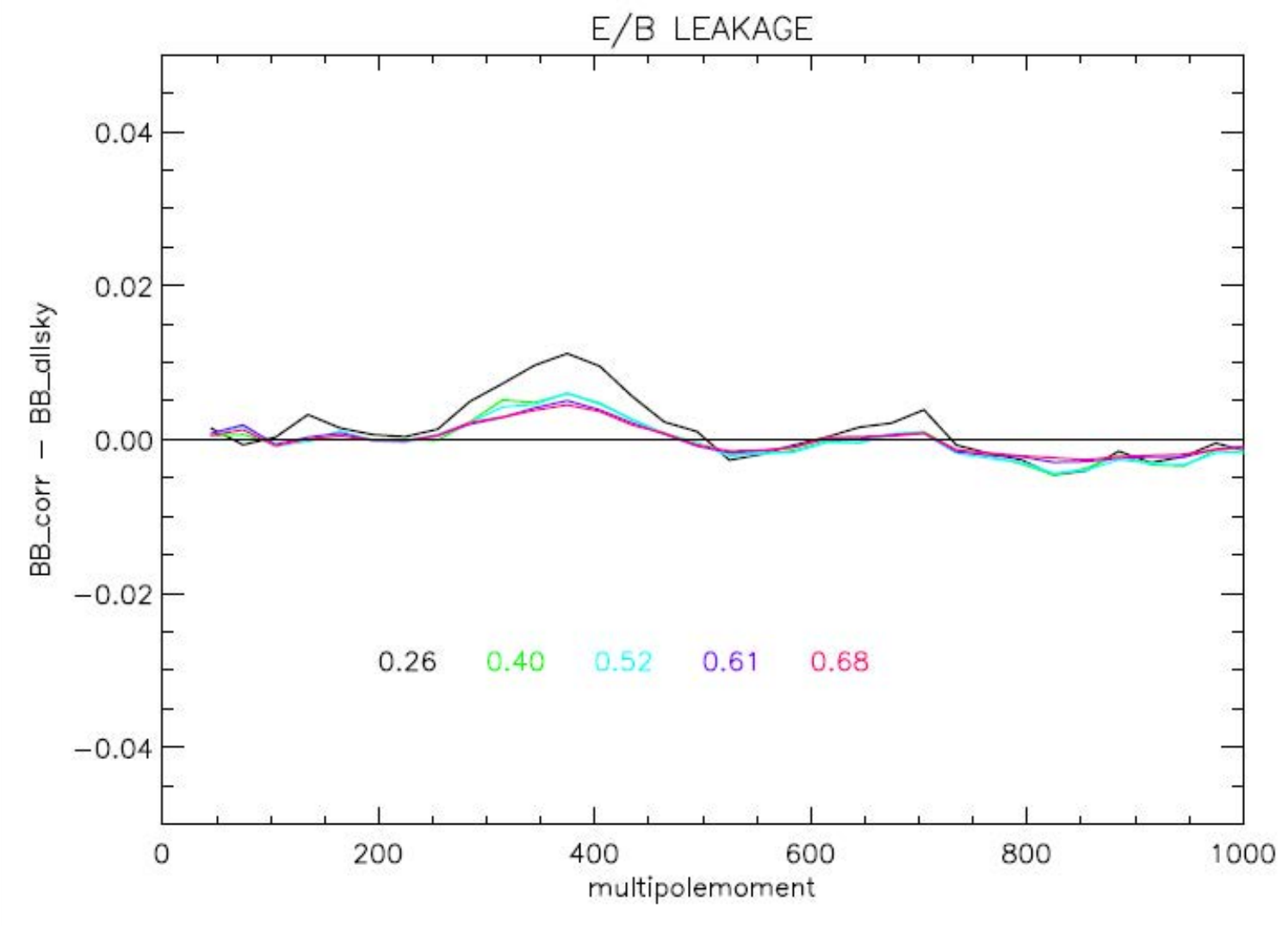}
\caption{The BB(mask) corrected - BB(full sky), for the CAMB T/S = 1 model. It is seen that the E/B leakage due to the applied mask is much smaller than the detected BB power spectrum. Y - axis: $l(l +1)/(2\pi) C^{EB}_{l}[\mu K^{2}]$}
\label{fig_9a}
\end{figure}

\begin{figure}[!]
\centering
\includegraphics[width=3.0 in]{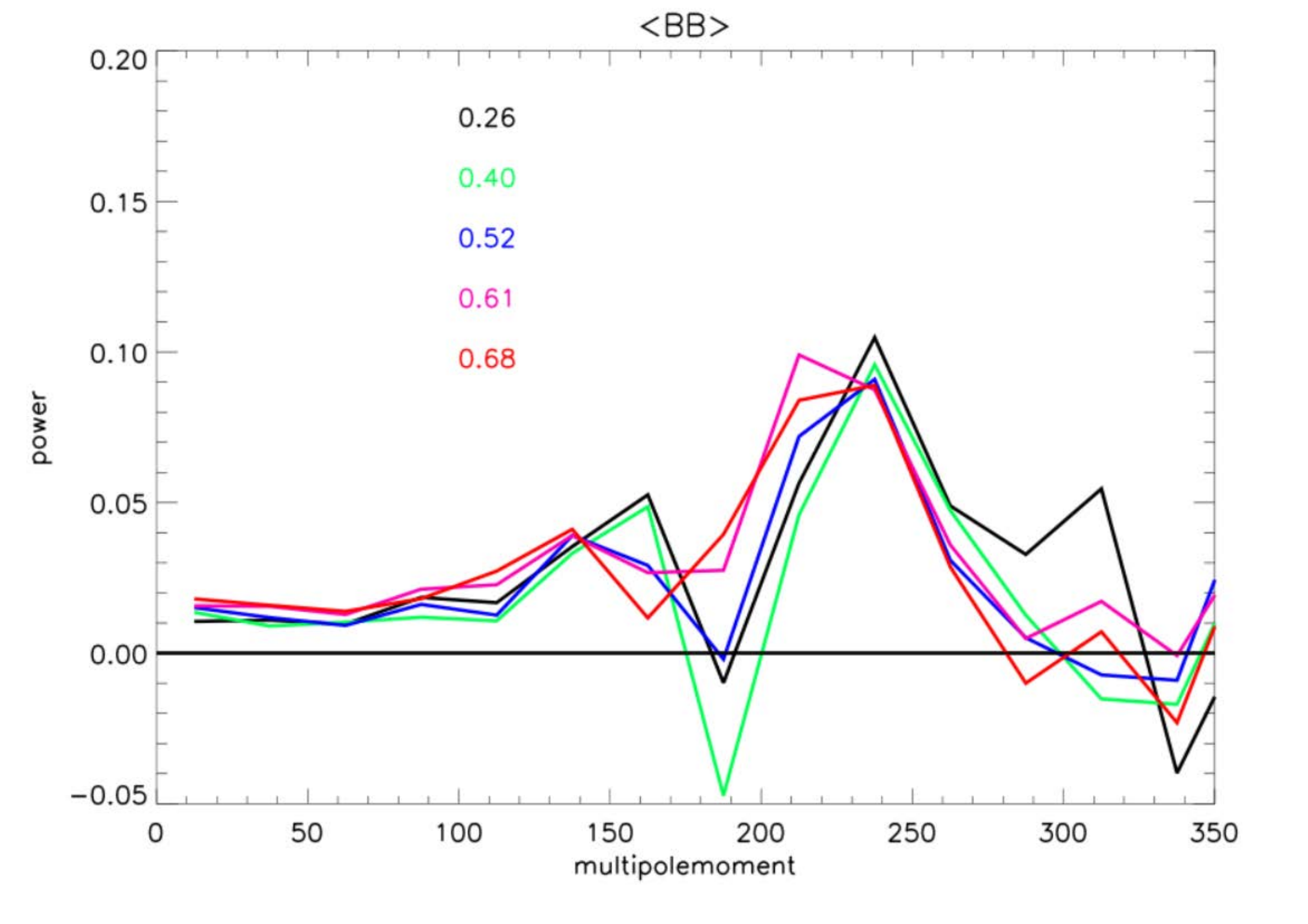}
\caption{The $<$HM, OE$>$ CC BB power spectra obtained with a set of CAMSPEC masks, with sky covering factors of 0.26, 0.40, 0.52, 0.61, 0.68. The spectra have been normalized to the covering factor of the mask used in obtained the HM and ODd/EVEN spectra. If the accuracy of the spectra is taken into account thee are no trend as function of the sky coverage. Y - axis: $l(l +1)/(2\pi) C^{BB}_{l}[\mu K^{2}]$}
\label{fig_9}
\end{figure}

A simple test to estimate the contribution of the Galactic components is to investigate how the cross power spectra are depending of the Galactic latitude.
In Fig. \ref{fig_9} the BB power spectra obtained within different CAMSPEC masks with different sky coverage (0.26 - 0. 68) are shown. It is seen that within the accuracy there is no variation of the spectra as function of the sky coverage, indicating a non - significant Galactic contribution.

\begin{figure}[!]
\centering
\includegraphics[width=3.0 in]{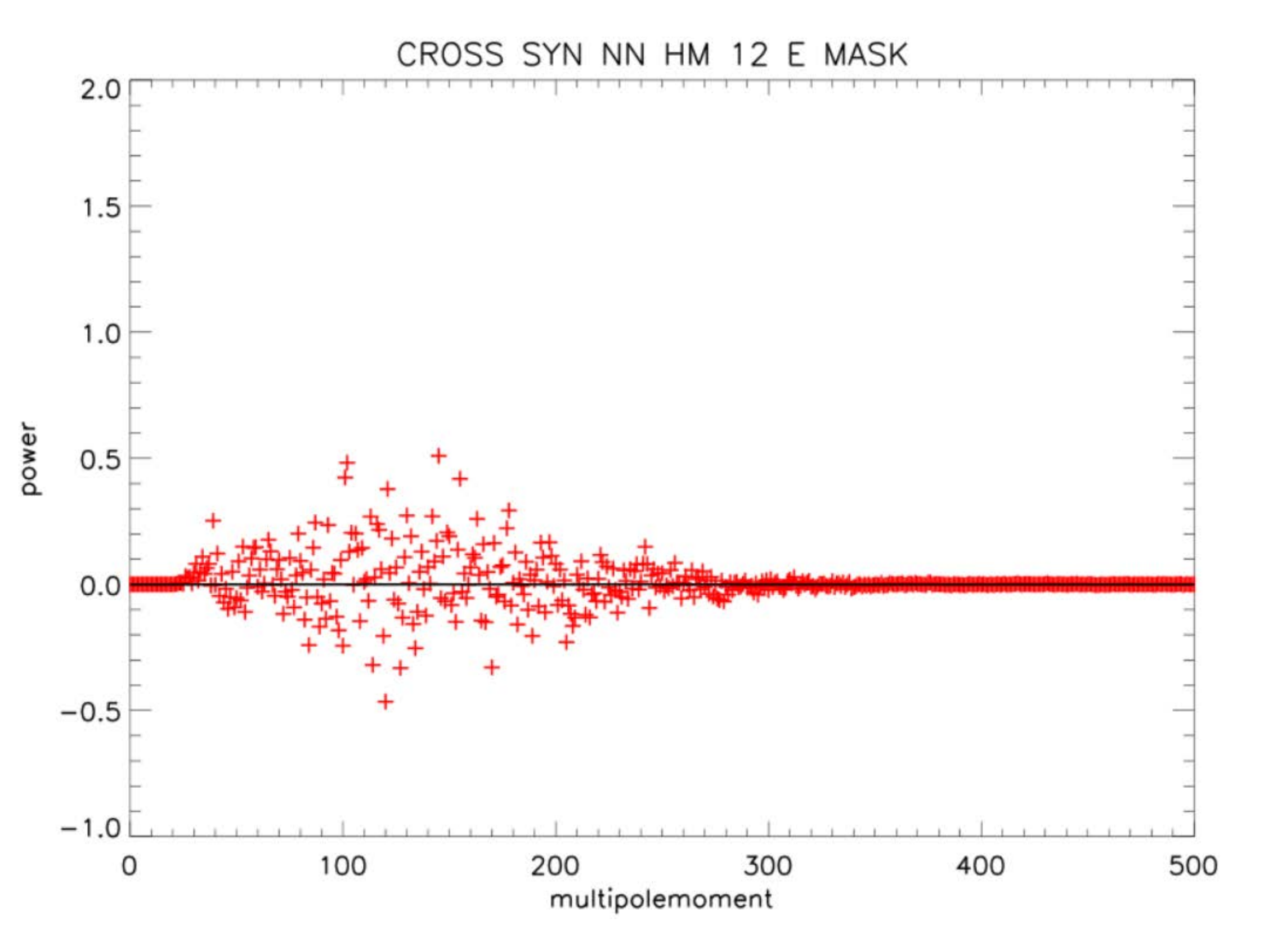}
\caption{The CC synchrotron HM 1 and NN HM 2 EE power spectrum. The synchrotron HM 1 map is provided by the Planck Commander team. Due to the observational errors in the LFI channels, they have applied a low pass filter removing multipolemoments above 250. It is evident that there is no evidence for a contamination of the NN maps by synchrotron emission from the Milky Way. X - axis: $\delta$l = 1, Y - axis: $l(l +1)/(2\pi) C^{EE}_{l}[\mu K^{2}]$}
\label{fig_10}
\end{figure}

\begin{figure}[!]
\centering
\includegraphics[width=3.0 in]{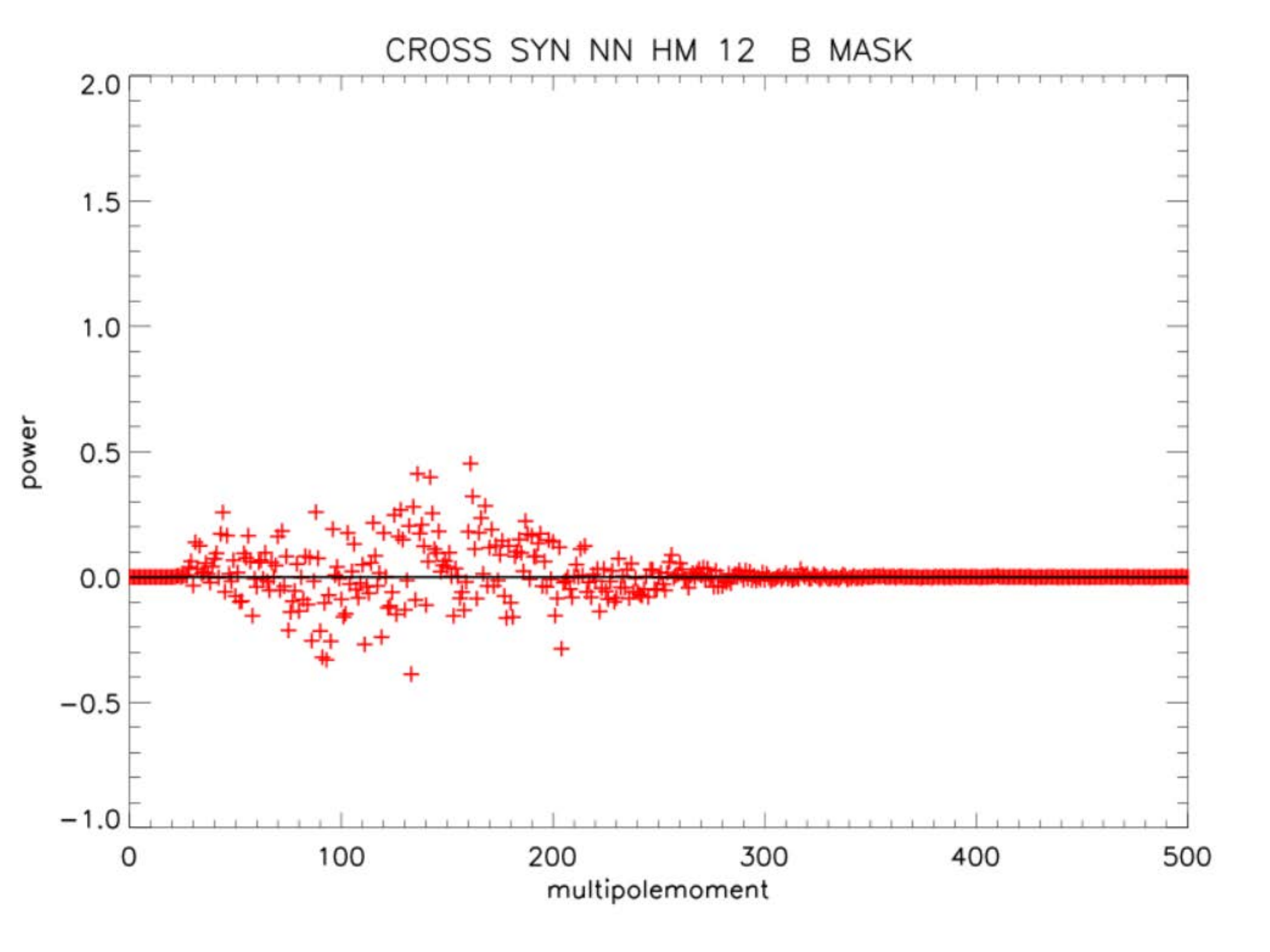}
\caption{The CC synchrotron HM 1 and NN HM 2 BB power spectrum. The synchrotron HM 1 map is provided by the Planck Commander team. Due to the observational errors in the LFI channels, they have applied a low pass filter removing multipolemoments above 250. It is evident that there is no evidence for a contamination of the NN maps by synchrotron emission from the Milky Way. X - axis: $\delta$l = 1,Y - axis: $l(l +1)/(2\pi) C^{BB}_{l}[\mu K^{2}]$}
\label{fig_11}
\end{figure}

\subsection{The synchrotron emission}

 It is generally accepted that only the Galactic synchrotron emission and the thermal dust emission are polarized. The best available maps of these components have been derived by the Commander team from the Planck 2017 data set. Due to the uncertainties in the LFI channels they have a applied a low pass filter.
In Figs. \ref{fig_10} and \ref{fig_11}, the CC EE and BB power spectra between the Commander synchrotron maps and the corresponding NN maps are shown.  It is evident that there is negligible synchrotron contribution to
the derived NN power spectra.

\begin{figure}[h]
\centering
\includegraphics[width=3.0 in]{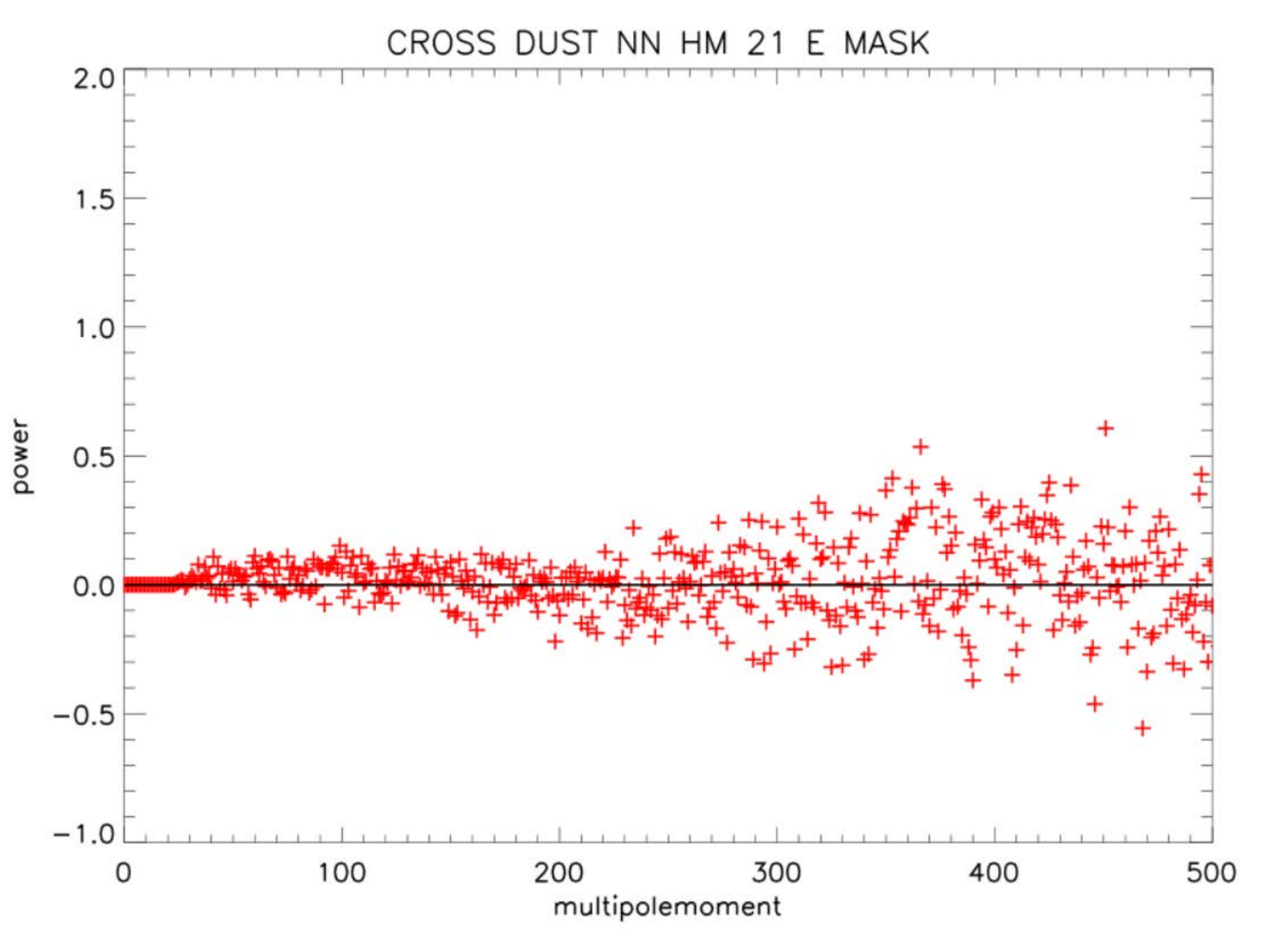}
\caption{The CC dust HM 1 and NN HM 2 EE power spectrum. The dust HM 1 map is provided by the Planck Commander team. It is evident that there is no sign of a contamination of the NN maps by thermal dust emission from the Milky Way. X - axis: $\delta$l = 1,Y - axis: $l(l +1)/(2\pi) C^{EE}_{l}[\mu K^{2}]$}
\label{fig_12}
\end{figure}

\begin{figure}[h]
\centering
\includegraphics[width=3.0 in]{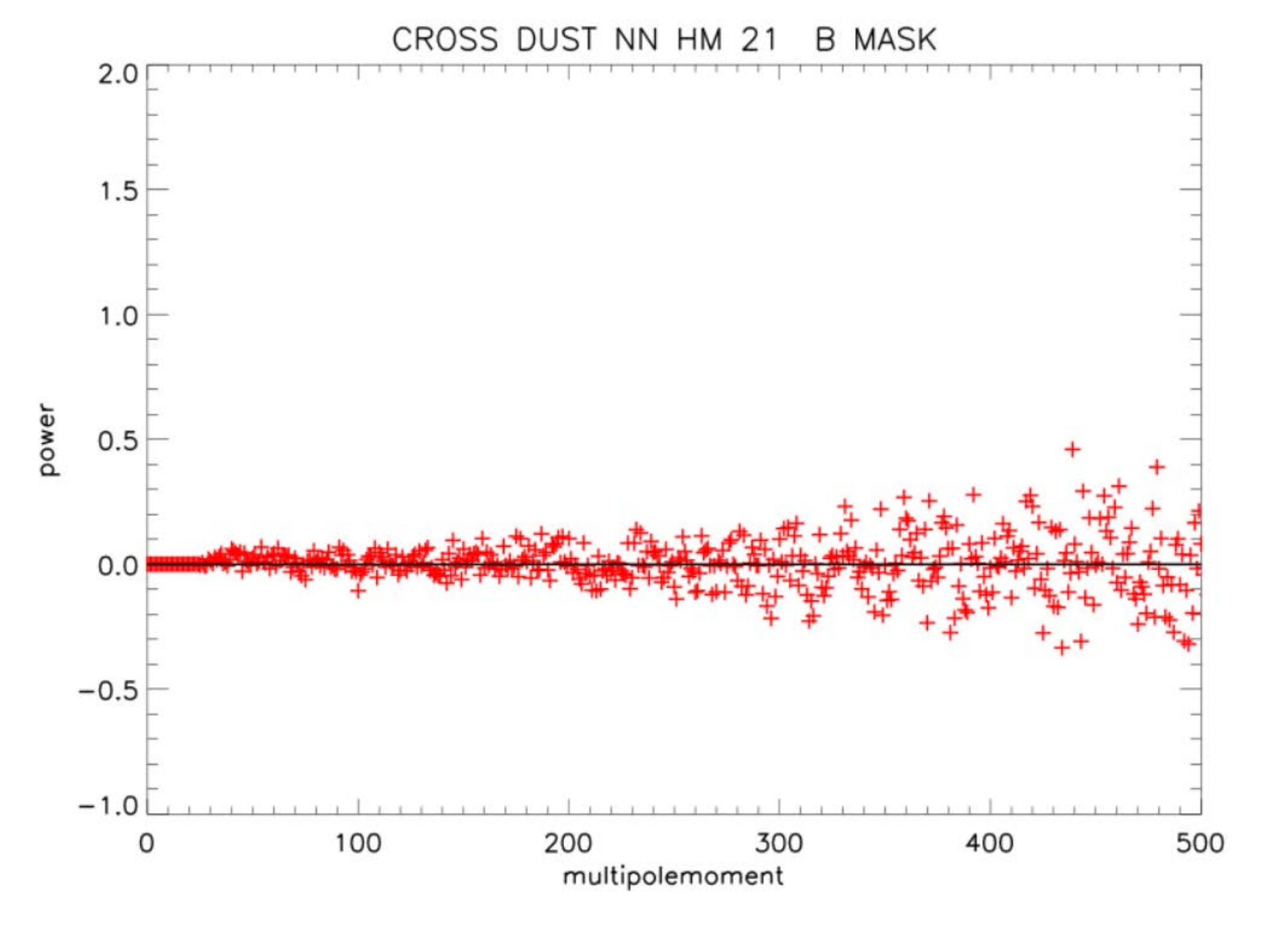}
\caption{The CC dust HM 1 and NN HM 2 BB power spectrum. The dust HM 1 map is provided by the Planck Commander team. It is evident that there is no evidence for a contamination of the NN maps by thermal dust emission from the Milky Way. X - axis: $\delta$l = 1,Y - axis: $l(l +1)/(2\pi) C^{BB}_{l}[\mu K^{2}]$}
\label{fig_13}
\end{figure}

\subsection{The dust thermal emission}

Similarly, in Figs. \ref{fig_12} and \ref{fig_13}, the CC EE and BB power spectra between the Commander thermal dust emission maps and the corresponding NN maps are shown. It is evident that there is also negligible contribution from dust to the derived NN power spectra.

\section{Conclusions}

It has been demonstrated that with the improved accuracy of the final Planck polarization maps compared the 2015 release, the detection of the BB power spectrum in NN1 is confirmed, with a higher confidence. Possible contamination from Galactic emission and remaining systematic errors in the Planck frequency maps has been ruled out. Two features in the EB power spectrum have been found each detected with a S/N $\sim$4 . At this stage, the most likely origin of these features is leakage from E modes to B modes of the order of 1 per cent. This level of leakage gives no significant contribution to the detected BB power spectrum. The TB power spectrum is found to be consistent with a zero spectrum.

The confirmation of the BB power spectrum will, no doubt, give new strong arguments for the proposed polarization missions to follow up on Planck.

\section{Acknowledgements}

The author acknowledges that this work would not have been possible without the massive efforts by a lot of strongly committed scientists and engineers within the Planck Collaboration. A anonymous referee is acknowledged for valuable comments. This work has taken advantage of the HEAlPix software package.



\end{document}